# New Developments in Time-Independent Quantum-Mechanical Perturbation Theory

A Kerley Technical Services Research Report

Gerald I. Kerley, Consultant

April 2008

A narrated PPT tutorial on this subject can be found at
http://kerleytechnical.com/tutorials.htm

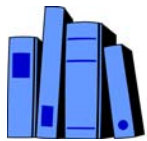

# New Developments in Time-Independent Quantum-Mechanical Perturbation Theory

A Kerley Technical Services Research Report

Gerald I. Kerley, Consultant

April 2008

## ABSTRACT

This report discusses two new ideas for using perturbation methods to solve the time-independent Schrödinger equation. The first concept begins with rewriting the perturbation equations in a form that is closely related to matrix diagonalization methods. That allows a simple, compact derivation of the standard (Rayleigh-Schrödinger) equations. But it also leads to a new iterative solution method that is not based on the usual power series expansion. The iterative method can also be used when the unperturbed system is two-fold degenerate. The second concept is quite different from the first but compatible with it. It is based on the fact that one can replace the true Hamiltonian with a synthetic Hamiltonian having the same eigenvalues. This approach allows one to cancel part of the perturbation and to reduce the size of the off-diagonal matrix elements, giving better convergence of the series. These methods are illustrated by application to three perturbed harmonic oscillator problems—a 1-D oscillator with a linear perturbation, a 1-D oscillator with a quartic perturbation, and a perturbed 2-D oscillator, which involves degenerate states. The iterative method and the synthetic Hamiltonian method are shown to give significantly better results than the standard method.

# PREFACE

Several years ago, I began a review of topics in quantum mechanics that are important for equation of state theory and related applications. While studying these issues, I decided that improvements to standard perturbation techniques could be helpful in addressing some of the unsolved problems.

This report discusses the new ideas that came out of my work. I have not been able to find any discussion of these ideas in the quantum mechanics or mathematics literature—except for my own 1974 paper on electron correlation. I would be grateful to know of any previous work that employs the same or similar methods.

I welcome constructive comments on this work, especially ideas for improving the methods and applying them to specific problems.

Gerald I. Kerley
Appomattox, VA
April 2008

I recently discovered that the technique used in my 1974 paper on electron correlation in the homogeneous electron gas was a variation of a method originally proposed by Boys and Handy in several papers written in 1969. Their approach, which is called the transcorrelated (TC) method, also used a similarity transformation of the Hamiltonian to treat electron correlation in atoms and molecules. I did not learn of their work until last year, after completing the present report.

It appears that the TC method has had little impact on quantum mechanical calculations until recently. In the last ten years, there has been a revival of interest in this approach, especially among researchers in Japan. For example, see the recent paper by S. Tsuneyuki, "Transcorrelated Method: Another Possible Way towards Electronic Structure Calculation of Solids," Progress of Theoretical Physics Supplement No. 176 (2008) pp. 134-142. I thank Professor Tsuneyuki for sending me information on this work.

The work discussed in this report is not limited to the electron correlation problem. The TC method is only one application of what I call "synthetic" Hamiltonians, in which a non-unitary similarity transformation is treated as a tool in perturbation theory.

Gerald I. Kerley
Appomattox, VA
July 2009

# CONTENTS

# FIGURES AND TABLES

# 1. INTRODUCTION

## 1.1 Background

In 1926, E. Schrödinger published a classic series of papers [1], presenting the equation that bears his name and applying it to several problems in quantum mechanics. One of these papers introduced time-independent perturbation theory and applied it to the calculation of the Stark effect in the hydrogen atom. This method, a standard topic in quantum mechanics texts [2][3], is often called "Rayleigh-Schrödinger perturbation theory" (RSPT) [4]. It is an important tool for obtaining solutions to the Schrödinger equation for complicated systems.

In perturbation theory, the wavefunctions and energies for the system of interest are expanded in terms of those for an "unperturbed" system having known solutions. The Hamiltonian matrix, when computed in terms of the unperturbed basis functions, contains diagonal and off-diagonal elements that are treated as perturbations. In RSPT, the corrections are expressed in powers of the perturbation matrix elements. Recursion relations express the wavefunction and energy corrections for a given order in terms of the corrections for all lower orders.

Despite its elegance and simplicity, RSPT has well-known difficulties that limit its usefulness in a number of important problems. The expansion does not converge unless the perturbation matrix elements between the unperturbed states are small compared to the energy differences between those states. In cases of degeneracy, where two or more unperturbed states have the same energy, zeros appear in the denominators and the method fails completely.

Various techniques have been developed to circumvent these problems. In cases of degeneracy, the standard recommendation is to take linear combinations of the unperturbed wavefunctions such that the perturbation matrix elements vanish [2][3]. This condition is achieved by diagonalizing the submatrix containing the degenerate states. A very different approach is to sum selected terms in the perturbation expansion, usually identified using diagrammatic methods, to all orders [5]. This approach is, in effect, a reordering of the terms in the series.

The success of these alternate methods demonstrates an important point. The divergences in RSPT do not necessarily indicate deficiencies in modeling the physics of the problem, e.g., a poor choice for the unperturbed system. Rather, the divergences appear because the RSPT power series orders the terms in an artificial way. Matrix diagonalization methods, which do not use any ordering scheme, can be used to obtain solutions even when the perturbation term is large. Alternately, diagrammatic methods can be used to reorder the terms and eliminate the divergences introduced by the RSPT power series.

Matrix methods and reordering schemes are most helpful when they are used as computational tools and applied to specific problems. By contrast, perturbation theory can be derived for a general case and is also amenable to some analytic manipulation. But perturbation theory would be more useful if it could incorporate some of the power of the matrix methods and reordering schemes while retaining this generality.[1]

## 1.2 Overview of Report

This report presents two new methods for improving time-independent perturbation theory. I will show that RSPT is only one of many recursive relationships for computing the wavefunctions and energies of perturbed systems and discuss an alternate scheme that gives better convergence than RSPT. In addition, I will show that the true Hamiltonian can be replaced by a "synthetic" Hamiltonian that has the same eigenvalues but reorders the perturbation corrections.[2]

The remainder of this report is organized as follows.

- Section 2 discusses the basic equations and notation used in the report.

- Section 3 derives RSPT from the matrix equations of Sec. 2. It also discusses an improved treatment of the diagonal matrix elements in RSPT.

- Section 4 discusses a new iterative method for solving the perturbation equations that gives better results than RSPT.

- Section 5 discusses synthetic Hamiltonians and how they can be used to reorder the perturbation corrections.

- Sections 6, 7, and 8 apply the new methods to three examples—the 1-D harmonic oscillator with a linear perturbation and a quartic perturbation, respectively, and a perturbed 2-D oscillator with degenerate unperturbed states.

- Section 9 gives a summary and conclusions.

- Appendices A and B give listings of FORTRAN subroutines for RSPT and the iterative perturbation method.

- Appendix C discusses the calculation of matrix elements for the harmonic oscillator problems.

---

1. Formulations of perturbation theory in terms of matrix methods already exist in the literature [6]-[8]. However, those papers employ matrix methods for pedagogical and/or computational purposes. The goals of this report are quite different.
2. I previously used a synthetic Hamiltonian in a study of electron correlation in the homogeneous electron gas [9]. The present work generalizes and extents those concepts.

                                                                   

## 2. BASIC EQUATIONS AND NOTATION

We want to solve the Schrödinger equation,

$$\mathcal{H}\,\Psi_k \,=\, E_k\Psi_k, \tag{1}$$

where $\mathcal{H}$ is the Hamiltonian operator for some system, and $\Psi_k$ and $E_k$ are the wavefunctions and energies (eigenfunctions and eigenvalues of $\mathcal{H}$).

Begin with the usual assumption that the $\Psi_k$ can be written as a linear combination of wavefunctions $\Phi_k$ for some "unperturbed system,"

$$\Psi_k \,=\, \Sigma_j\,\Phi_j U_{jk}, \tag{2}$$

assuming that the $\Phi_k$ are a complete set. I will also make the usual requirement of orthonormality for the *unperturbed* wavefunctions,

$$\int \Phi_k^*\Phi_l\,d\tau \,=\, \delta_{kl}. \tag{3}$$

(The orthogonality and normalization of the perturbed wavefunctions will remain unspecified for now.)

In standard treatments of perturbation theory, the $\Phi_k$ are usually taken to be eigenfunctions of an "unperturbed Hamiltonian" $\mathcal{H}_0$. However, there are cases where it is inconvenient to define $\mathcal{H}_0$, and I will demonstrate that such a definition is not actually necessary.

Substitute Eq. (2) into Eq. (1), multiply by $\Phi_l^*$, integrate, and use Eq. (3). Then

$$\Sigma_j\,H_{lj}U_{jk} \,=\, E_k U_{lk}, \text{ where } H_{lj} \,=\, \int \Phi_l^*\mathcal{H}\,\Phi_j\,d\tau \equiv \langle l\,|\,\mathcal{H}\,|\,j\rangle. \tag{4}$$

This set of equations can also be written

$$\Sigma_j\,H_{lj}C_{jk} \,=\, E_k C_{lk}, \tag{5}$$

$$E_k \,=\, \Sigma_j\,H_{kj}C_{jk}, \text{ where} \tag{6}$$

$$C_{lk} \,=\, U_{lk}/U_{kk}, \text{ and } C_{kk} \,=\, 1\,. \tag{7}$$

RSPT and other perturbation theories can be derived from Eqs. (5), (6) and (7).

Some additional remarks are needed before leaving this section.

First, I have deliberately avoided the assumption that $\mathcal{H}$ is a Hermitian operator. Hence the above equations, and those of the next two sections, will also be applicable to problems involving synthetic Hamiltonians (Sec. 5), which are *not* Hermitian.

Second, it is useful to note the equivalence between the above equations and those for matrix diagonalization. Using matrix notation, Eq. (4) can be written

$$\boldsymbol{HU} = \boldsymbol{UE}, \tag{8}$$

where $\boldsymbol{E}$ is a diagonal matrix ($[\boldsymbol{E}]_{kl} = E_k \delta_{kl}$). Multiplying both sides of Eq. (8) with the inverse of $\boldsymbol{U}$, one obtains

$$\boldsymbol{U}^{-1}\boldsymbol{HU} = \boldsymbol{E}. \tag{9}$$

Hence the matrix $\boldsymbol{U}$ diagonalizes the Hamiltonian matrix $\boldsymbol{H}$. Since Eqs. (5)-(7) determine the matrices $\boldsymbol{E}$ and $\boldsymbol{U}$ (except for normalization of the perturbed wavefunctions), perturbation equations derived from these equations are equivalent to those for matrix diagonalization.

Once again, the matrix equations are valid whether or not $\boldsymbol{H}$ is Hermitian. But, if it *is* Hermitian, it is well-known that $\boldsymbol{U}^{-1} = \boldsymbol{U}^T$, so that

$$\boldsymbol{U}^T\boldsymbol{HU} = \boldsymbol{E}, \ \boldsymbol{U}^T\boldsymbol{U} = \boldsymbol{I} \ (\text{if } \boldsymbol{H} \text{ is Hermitian}). \tag{10}$$

Most of the ideas in this report were developed to deal with cases where the matrices are of infinite extent. That condition is especially important in Sec. 5, where certain operator relations require summations over the complete set of functions $\Phi_k$. However, the methods discussed in Secs. 3 and 4 can also be used for finite matrices, i.e., where a subset of the full Hamiltonian matrix is used in an approximate calculation.

# 3. RALEIGH-SCHRÖDINGER THEORY WITH IMPROVEMENTS

In this section I will present a compact derivation the RSPT equations, using Eqs. (5)-(7). I will also discuss two new results that could be useful in applications of the theory.

As usual, start by introducing an expansion (ordering) parameter $\lambda$ that will later be set to unity. Then write the Hamiltonian matrix elements in the form

$$H_{lk} = E_k^0 \delta_{lk} + \lambda W_{lk}. \tag{11}$$

Note that I have not defined an "unperturbed Hamiltonian" $\mathcal{H}_0$ or even required the energies $E_k^0$ to be eigenvalues of any operator. Therefore, one has some freedom in defining these energies, as discussed below.

Next write the expansion coefficients $C_{lk}$ as a power series in $\lambda$.

$$C_{lk} = \Sigma_{\alpha=0}^{\infty} \lambda^{\alpha} C_{lk}^{(\alpha)}, \tag{12}$$

and require that $C_{lk}^{(0)} = \delta_{lk}$, $C_{kk}^{(\alpha)} = 0$ for $\alpha \geq 1$. Using Eq. (6), it follows that

$$E_k = E_k^0 + \Sigma_{\alpha=1}^{\infty} \lambda^{\alpha} E_k^{(\alpha)}, \text{ where } E_k^{(\alpha)} = \Sigma_l W_{kl} C_{lk}^{(\alpha-1)}. \tag{13}$$

Inserting Eqs. (11)-(13) into Eq. (5), and canceling some terms, one obtains

$$\Sigma_{\alpha=1}^{\infty} \lambda^{\alpha} [(E_l^0 - E_k^0) C_{lk}^{(\alpha)} + \Sigma_j W_{lj} C_{jk}^{(\alpha-1)}] - \Sigma_{\alpha=1}^{\infty} \Sigma_{\beta=0}^{\infty} \lambda^{\alpha+\beta} E_k^{(\alpha)} C_{lk}^{(\beta)} = 0. \tag{14}$$

The first-order equations, obtained by collecting all terms having $\lambda$ to the first power, are

$$(E_l^0 - E_k^0) C_{lk}^{(1)} + \Sigma_j W_{lj} C_{jk}^{(0)} - E_k^{(1)} C_{lk}^{(0)} = 0. \tag{15}$$

Hence

$$E_k^{(1)} = W_{kk}, \tag{16}$$

$$C_{lk}^{(1)} = W_{lk} / (E_k^0 - E_l^0), \, l \neq k. \tag{17}$$

Using Eqs. (13) and (17), the second-order energy expression is given by

$$E_k^{(2)} = \Sigma_{l \neq k} W_{kl} W_{lk} / (E_k^0 - E_l^0). \tag{18}$$

For $\alpha \geq 2$, Eq. (14) gives the following recursion relation for the expansion coefficients.

$$C_{lk}^{(\alpha)} = (E_k^0 - E_l^0)^{-1} [\Sigma_{j \neq k} W_{lj} C_{jk}^{(\alpha-1)} - \Sigma_{\beta=1}^{\alpha-1} C_{lk}^{(\alpha-\beta)} E_k^{(\beta)}], \ l \neq k. \tag{19}$$

The corresponding energy corrections are given by Eq. (13).

The above relations, which allow calculation of the perturbation corrections to an arbitrarily high order, are identical to the ones derived by the usual methods [4]. However, the present derivation leads to two new results.

First, these formulas have been obtained without specifying an unperturbed Hamiltonian $\mathcal{H}_0$ or perturbation $W = \mathcal{H} - \mathcal{H}_0$. In order to apply the equations, one needs only to compute the matrix elements of the full Hamiltonian in the chosen "unperturbed" basis set, just as is done in matrix diagonalization. This result can be useful in cases where it is inconvenient to define $\mathcal{H}_0$ and $W$ for the system.

Second, and even more important, it is not necessary to assume that the zeroth-order energies $E_k^0$ are eigenvalues of a Hamiltonian $\mathcal{H}_0$. In fact, there is some freedom in defining these energies. One can even set $E_k^0 = H_{kk}$, if desired, so that $W_{kk} = 0$ and the perturbation terms $W_{lk}$ only involve the off-diagonal matrix elements of the Hamiltonian.

Note that the standard RSPT formulas still apply if one chooses $E_k^0 = H_{kk}$, except that the zeroth-order energy differences $E_k^0 - E_l^0$ are replaced by the first-order energy differences $H_{kk} - H_{ll}$. This modification could be useful in cases where the first-order energy correction breaks the degeneracy of the unperturbed system.[1]

Appendix A lists a FORTRAN subroutine that calculates the eigenvalues and eigenvectors of a Hamiltonian matrix using the RSPT method, the diagonal elements being treated in the manner discussed above. This subroutine was used in some of the calculations discussed in Secs. 6 and 7.

Before leaving this section, I reiterate the point, made in Sec. 2, that the above formulas do *not* require the Hamiltonian operator $\mathcal{H}$ to be Hermitian.

---

1. Of course, there are many cases where the first-order correction *does* not break the degeneracy. In such cases, a different choice of the $E_k^0$ could still allow a solution of the degeneracy problem.

## 4. AN ITERATIVE METHOD

Equations (5)-(7), derived in Sec. 2, express the wavefunctions and energies of a perturbed system in terms of those for an unperturbed system. All perturbation theories represent attempts to solve these equations. RSPT, derived from the power series expansion in Sec. 3, is *one* method of solution but not necessarily the *best* one.

In this section, I will discuss an *iterative* method for solving the perturbation equations: One starts with an approximate solution, then uses the equations to find a better estimate of the solution. The new approximation is used to obtain a still better estimate, and the process is continued until convergence is obtained.

There are several ways of writing Eqs. (5)-(7) that are suitable for iteration. In this section, I will discuss one method that I have found to give good results for some problems.

Combining Eqs. (5) and (6) gives

$$\Sigma_j \, H_{lj} C_{jk} - C_{lk} \, \Sigma_j \, H_{kj} C_{jk} \,=\, 0 \,. \tag{20}$$

It is convenient to define the quantities

$$\Delta_{kl} \,=\, H_{kk} - H_{ll} \,, \tag{21}$$

$$W_{kl} \,=\, H_{kl} \text{ for } k \neq l \text{, and } W_{kk} \,=\, 0 \,. \tag{22}$$

Using these definitions, and noting that $C_{kk} \,=\, 1$ [Eq. (7)], Eq. (20) can be written

$$W_{kl} C_{lk}^2 + \Delta_{kl} C_{lk} - Y_{lk} \,=\, 0 \,, \tag{23}$$

$$Y_{lk} \,=\, W_{lk} + \Sigma_{j \neq k, l} \, (W_{lj} - C_{lk} W_{kj}) C_{jk} \,. \tag{24}$$

Treating Eq. (23) as a quadratic equation, one finds

$$C_{lk} \,=\, \frac{-\Delta_{kl} \pm \sqrt{\Delta_{kl}^2 + 4 W_{kl} Y_{lk}}}{2 W_{kl}} \,=\, \frac{2 Y_{lk}}{\Delta_{kl} \pm \sqrt{\Delta_{kl}^2 + 4 W_{kl} Y_{lk}}} \,. \tag{25}$$

The sign in Eq. (25) can be chosen by comparison with second-order RSPT. Define

$$s_{kl} = 1 \text{ for } \Delta_{kl} > 0 \text{ and } s_{kl} = -1 \text{ for } \Delta_{kl} < 0. \tag{26}$$

Now require that Eq. (25) give the same result for the energy as Eq. (18), in the limit that the matrix elements $W_{kl}$ are small compared with the energy differences $\Delta_{kl}$. Then Eq. (25) becomes

$$C_{lk} = 2 s_{kl} Y_{lk} / (\sqrt{\Delta_{kl}^2 + 4 W_{kl} Y_{lk}} + |\Delta_{kl}|). \tag{27}$$

It is useful to generalize this result to the case $\Delta_{kl} = 0$, in which the sign can be chosen arbitrarily. I will use the following procedure: If $\Delta_{kl} \neq 0$, one can arrange the matrix elements such that the diagonal elements $H_{kk}$ increase with index $k$. Thus $s_{kl}$ has the same sign as $k - l$. Retaining this relation as $\Delta_{kl} \to 0$, one obtains

$$s_{kl} = sign(k - l) \text{ for } \Delta_{kl} = 0. \tag{28}$$

Equation (27) is not yet a complete solution to the problem, because the coefficient matrix elements $C_{lk}$ also appear in the quantities $Y_{lk}$. However, it does offer the possibility of an iterative solution: Start with guesses for the $C_{lk}$, compute the $Y_{lk}$, then use Eq. (27) to obtain new values for the $C_{lk}$ and Eq. (6) to compute the energies. This process is then repeated until satisfactory convergence is obtained.

Let $C_{lk}^{(\mu)}$ and $E_k^{(\mu)}$ denote the coefficients and energies obtained after $\mu$ iterations,

$$C_{lk}^{(\mu)} = 2 s_{kl} Y_{lk}^{(\mu-1)} / (\sqrt{\Delta_{kl}^2 + 4 W_{kl} Y_{lk}^{(\mu-1)}} + |\Delta_{kl}|), \tag{29}$$

$$E_k^{(\mu)} = H_{kk} + \Sigma_{l \neq k} W_{kl} C_{lk}^{(\mu)} = H_{kk} + \varepsilon_k^{(\mu)}. \tag{30}$$

Now begin the iteration with $C_{lk}^0 = \delta_{lk}$, $Y_{lk}^0 = W_{lk}$. The first iteration gives

$$C_{lk}^{(1)} = 2 s_{kl} W_{lk} / (\sqrt{\Delta_{kl}^2 + 4 W_{kl} W_{lk}} + |\Delta_{kl}|), \tag{31}$$

$$\varepsilon_k^{(1)} = \Sigma_{l \neq k} 2 s_{kl} W_{kl} W_{lk} / (\sqrt{\Delta_{kl}^2 + 4 W_{kl} W_{lk}} + |\Delta_{kl}|). \tag{32}$$

These expressions are closely related to those obtained in RSPT. If $W_{kl}$ is small compared to $\Delta_{kl}$, the denominator can be expanded as follows.

$$\sqrt{\Delta_{kl}^2 + 4 W_{kl} W_{lk}} + \left| \Delta_{kl} \right| = 2 \left| \Delta_{kl} \right| + 2 W_{kl} W_{lk} / \left| \Delta_{kl} \right| + \dots \approx 2 \left| \Delta_{kl} \right|. \qquad (33)$$

Substituting the above approximation into Eqs. (31) and (32) reproduces Eqs. (17) and (18) of RSPT.

However, these new relations differ from those of RSPT in a very important way—they do not "blow up" when $\Delta_{kl} \to 0$. Moreover, they give the *exact* solution to diagonalization of a 2 x 2 matrix, even in the degenerate case.[1]

These formulas obviously become more complicated when the iteration is carried to higher order, but the calculations are not intractable, particularly when numerical methods are used. In fact, this method is quite suitable for numerical computation, because the algorithm only needs to save the results of the current and previous iterations. (In RSPT, computation of a given order correction requires the results of all lower orders.)

This method does have certain limitations that should be noted.

- The argument of the square root in Eq. (29) can become negative after two or more iterations. In this report, I take the position that this situation represents a failure of the algorithm to converge.

- The method gives incomplete results when the diagonal matrix elements $H_{kk}$ are three-fold or higher degenerate. In that case, the iteration may converge for some eigenvalues but not for others.

- For three-fold and higher degeneracy, some (but not all) of the converged eigenvalues may be incorrect, as shown in Sec. 8. I have not yet identified the reason for this problem.

It may be possible to eliminate the above difficulties by changing the initial guess, by modifying the iteration step, or by changing the sign convention, Eq. (28).

Appendix B lists a FORTRAN subroutine that calculates the eigenvalues and eigenvectors of a Hamiltonian matrix using the iterative perturbation method. It is quite simple and straightforward. This subroutine was used in some of the calculations discussed in Secs. 6 and 7.

This subroutine is used with Hamiltonian matrices having finite dimensions. In certain cases, it may also be possible to apply the numerical iteration method to matrices of infinite dimensions. However, I have not examined that possibility in this report.

---

1. For example, compare with the results given in Sec. 6.2.1 of Griffith's book [3].

Before leaving this section, I want to note two important points. First, examination of Eqs. (23) and (24) shows that the value of a particular coefficient, $C_{lk}$, depends only on the other coefficients $C_{jk}$ corresponding to the same eigenvalue $E_k$. Therefore, *this solution procedure can be applied to each eigenstate of the system, one at a time*, just as in RSPT; it does not require complete solution for all eigenstates, in contrast to matrix diagonalization methods.

The second point follows from the first one. *There is a one-to-one correspondence between the initial unperturbed eigenstates and the final perturbed eigenstates of the Hamiltonian.*[1] Hence this procedure really is a perturbation method, not a matrix diagonalization method in disguise.

Finally, I reiterate that the above formulas do *not* require the Hamiltonian operator $\mathcal{H}$ to be Hermitian, as observed in Secs. 2 and 3.

---

1. As already noted, this correspondence is arbitrary in cases of degeneracy, depending upon how the unperturbed states are ordered in the Hamiltonian matrix.

# 5. SYNTHETIC HAMILTONIANS

In this section, I will show how the true Hamiltonian operator can be replaced by a "synthetic" Hamiltonian that has the same eigenvalues. I will then consider the properties of this new operator and its potential role in perturbation theory.

## 5.1 Matrix Equations

Consider the following non-unitary transformation of the Hamiltonian matrix. Given a real matrix $\boldsymbol{S}$, let

$$\widetilde{\boldsymbol{H}} = e^{\boldsymbol{S}} \boldsymbol{H} \, e^{-\boldsymbol{S}} \text{ or } \boldsymbol{H} = e^{-\boldsymbol{S}} \widetilde{\boldsymbol{H}} \, e^{\boldsymbol{S}} \,, \tag{34}$$

where the exponentials are defined by

$$e^{\pm\boldsymbol{S}} = \Sigma_{k=0}^{\infty} \, (\pm\boldsymbol{S})^k / k! \,, \text{ with } e^{\boldsymbol{S}} e^{-\boldsymbol{S}} = e^{-\boldsymbol{S}} e^{\boldsymbol{S}} = \boldsymbol{I}. \tag{35}$$

Since the true Hamiltonian matrix for the system must be Hermitian, it can be diagonalized by a unitary transformation, Eq. (10). But one can also write

$$\boldsymbol{E} = \boldsymbol{U}^T \boldsymbol{H} \boldsymbol{U} = \widetilde{\boldsymbol{U}}^{-1} \widetilde{\boldsymbol{H}} \, \widetilde{\boldsymbol{U}} \,, \text{ where } \widetilde{\boldsymbol{U}} = e^{\boldsymbol{S}} \boldsymbol{U}, \ \widetilde{\boldsymbol{U}}^{-1} = \boldsymbol{U}^T e^{-\boldsymbol{S}}. \tag{36}$$

Hence $\widetilde{\boldsymbol{H}}$ has the same eigenvalues as $\boldsymbol{H}$; one can substitute $\widetilde{\boldsymbol{H}}$ for $\boldsymbol{H}$ when calculating $\boldsymbol{E}$. And, since perturbation theory gives the same eigenvalues as matrix diagonalization, as shown in Sec. 2, it can also be used to compute the eigenvalues.

Note that $\widetilde{\boldsymbol{H}}$ is not Hermitian. Equation (34) can be written

$$\widetilde{\boldsymbol{H}} = \boldsymbol{H} + \boldsymbol{F} - \boldsymbol{G}, \tag{37}$$

$$\boldsymbol{F} = \tfrac{1}{2}(e^{\boldsymbol{S}} \boldsymbol{H} \, e^{-\boldsymbol{S}} - e^{-\boldsymbol{S}} \boldsymbol{H} \, e^{\boldsymbol{S}}), \tag{38}$$

$$\boldsymbol{G} = \boldsymbol{H} - \tfrac{1}{2}(e^{\boldsymbol{S}} \boldsymbol{H} \, e^{-\boldsymbol{S}} + e^{-\boldsymbol{S}} \boldsymbol{H} \, e^{\boldsymbol{S}}). \tag{39}$$

It is easily seen that $\boldsymbol{F}$ is an anti-Hermitian matrix, while $\boldsymbol{G}$ is Hermitian.

Note that $\widetilde{\boldsymbol{H}}$ actually represents a *family* of matrices, each member being generated by a particular choice of the transformation matrix $\boldsymbol{S}$. The object of the transformation is to configure a "synthetic" Hamiltonian that is more amenable to perturbation theory than the true Hamiltonian.

## 5.2 Operator Equations

The next task is to determine the operators corresponding to the matrices $\boldsymbol{S}$, $\widetilde{\boldsymbol{H}}$, $\boldsymbol{F}$, and $\boldsymbol{G}$. The Hamiltonian operator for a general many-body system is

$$\mathcal{H} = -\Sigma_i\,(\hbar^2/2m_i)\nabla_i^2 + V = \mathcal{K} + V, \tag{40}$$

where the sum is taken over all particles having masses $m_i$. An explicit expression for the kinetic energy operator $\mathcal{K}$ is needed for what follows. It is not necessary to give a specific expression for the potential energy $V$ at this point; however, I will assume that it consists only of multiplicative terms, e.g., that it does not depend on the particle momenta.

Assuming the matrices are infinite, i.e., summations are carried out over all eigenstates in the basis set $\Phi_k$, the matrix equations (34)-(39) can be replaced by operator equations. Hence Eq. (34) becomes

$$\widetilde{\mathcal{H}} = e^S \mathcal{H}\, e^{-S} = \mathcal{H} + e^S(\mathcal{H}\, e^{-S} - e^{-S}\mathcal{H}). \tag{41}$$

I will now restrict $S$ to be a multiplicative operator (like $V$). Then $S$ commutes with $V$, and the commutator in Eq. (41) becomes

$$e^S(\mathcal{H}\, e^{-S} - e^{-S}\mathcal{H}) = e^S(\mathcal{K}\, e^{-S} - e^{-S}\mathcal{K}) = \mathcal{F} - G, \tag{42}$$

where $\mathcal{F}$ and $G$ are the operators corresponding to the matrices $\boldsymbol{F}$ and $\boldsymbol{G}$. One obtains

$$\widetilde{\mathcal{H}} = \mathcal{H} + \mathcal{F} - G, \tag{43}$$

$$\mathcal{F} = \Sigma_i\,(\hbar^2/2m_i)(2\nabla_i S \cdot \nabla_i + \nabla_i^2 S) = S\mathcal{H} - \mathcal{H}\,S = S\mathcal{H}_0 - \mathcal{H}_0\,S,^1 \tag{44}$$

$$G = \Sigma_i\,(\hbar^2/2m_i)(\nabla_i S)^2 = \tfrac{1}{2}(\mathcal{F}\,S - S\mathcal{F}). \tag{45}$$

Note that $G$ is positive-definite. The corresponding matrix equations are

$$\boldsymbol{F} = \boldsymbol{SH} - \boldsymbol{HS} = \boldsymbol{SH_0} - \boldsymbol{H_0 S}, \tag{46}$$

$$\boldsymbol{G} = \tfrac{1}{2}(\boldsymbol{FS} - \boldsymbol{SF}). \tag{47}$$

---

1. One can replace $\mathcal{H}$ with some "unperturbed" Hamiltonian $\mathcal{H}_0$ in Eq. (44) because $S$ commutes with the potential energy. Hence the only terms contributing to the commutator come from the kinetic energy operator, which is the same in both cases.

Once again, note that these relations hold only if the matrices are infinite, i.e., in the limit that the summations are carried out over all eigenstates in the basis set $\Phi_k$. For finite matrices, the synthetic Hamiltonian does not have the same eigenvalues as the true Hamiltonian unless one uses Eqs. (39) and (40).

## 5.3  Application to Perturbation Theory

The matrix elements of $\widetilde{\mathcal{H}}$ are

$$\widetilde{H}_{lk} = H_{lk} + F_{lk} - G_{lk}. \tag{48}$$

Since $\mathcal{F}$ is anti-Hermitian, it does not contribute to the diagonal elements, and

$$\widetilde{H}_{kk} = H_{kk} - G_{kk} = \int \Phi_k^* (\mathcal{H} - G) \Phi_k \, d\tau \equiv \langle k \, | \mathcal{H} - G | \, k \rangle. \tag{49}$$

The above result shows that a synthetic Hamiltonian can be used to move some of the perturbation corrections to the diagonal, making them first-order.

In order to take advantage of this synthetic Hamiltonian option, one wants to choose a function $S$ so that the off-diagonal elements of $\widetilde{\mathcal{H}}$ are more suitable for perturbation theory than those of $\mathcal{H}$. This function can be chosen using both mathematical and physical arguments.

I will consider the mathematical arguments first. There are two strategies.

**Strategy #1.** Suppose the true Hamiltonian can be written $\mathcal{H} = \mathcal{H}_0 + W$, where the basis functions $\Phi_k$ are the known eigenfunctions of $\mathcal{H}_0$. Suppose one could find a function $S$ such that $G = W$, completely canceling the perturbation. If so, the off-diagonal elements of $\widetilde{H}_{lk}$ would all come from the operator $\mathcal{F}$. Using Eq. (46), the Hamiltonian matrix elements would be

$$H_{lk} = E_k^0 \delta_{lk} + (E_k^0 - E_l^0) S_{lk} = E_k^0 \delta_{lk} + \Delta_{kl} S_{lk}, \text{(if } G = W). \tag{50}$$

The off-diagonal terms would decrease in magnitude as $\Delta_{kl}$ decreases and vanish completely for degenerate eigenstates—a desirable result for perturbation theory.

It is not likely that one can satisfy the condition $G = W$ exactly in most problems. If that were possible, there would be no perturbation correction for degenerate states—an unlikely situation. Furthermore, since $G$ is a positive-definite function, it only be used to cancel positive terms in $W$. A more realistic strategy would be to find the function $G$ that cancels the more intractable parts of the perturbation $W$ or replaces them with simpler functions. (See the discussion in Sec. 5.4, below.)

**Strategy #2.** Since $\mathcal{F}$ is anti-Hermitian, $F_{kl} = -F_{lk}$, and it is impossible to zero out all of the off-diagonal matrix elements, Eq. (48). However, it is possible, in principle, to make $\boldsymbol{H}$ a *triangular* matrix, at least in simple cases. Since the diagonal elements of a triangular matrix are also its eigenvalues,[1] the energies could then be computed without using the perturbation equations. (These equations, or some other method, would still be needed to compute the wavefunctions.)

Once again, it may not be possible to satisfy this condition exactly in complicated problems, but an approximation to it should still give good results.

I will illustrate the use of Strategy #2 for a simple case, in Sec. 6.

**Different Hamiltonians for different eigenstates**. It may be that a particular choice for $S$ and $\mathcal{H}$ will give good results for some eigenstates but not for others. In Sec. 4, I pointed out that the new iterative solution algorithm can be carried out independently for each eigenstate—one does not need to obtain a global solution to the matrix. Therefore, one has the option of using a different synthetic Hamiltonian for each eigenstate.

## 5.4 The Wavefunctions

Although the eigenstates of $\mathcal{H}$ and $\widetilde{\mathcal{H}}$ have the same energies, they do not have the same *eigenfunctions*. The eigenfunctions of $\widetilde{\mathcal{H}}$ are

$$\widetilde{\Psi}_k = U_{kk} \Sigma_l \Phi_l C_{lk}, \tag{51}$$

where the $C_{lk}$ are determined from the perturbation equations and $U_{kk}$ is some normalization constant. (Since $\widetilde{\mathcal{H}}$ is not Hermitian, the transformation is not unitary, and the $\widetilde{\Psi}_k$ will not generally be orthogonal to one another.)

It is important to note that the $\widetilde{\Psi}_k$ are not eigenfunctions of $\mathcal{H}$ and so are not the true wavefunctions for the system. They should not be expected to give correct results when used to compute quantities other than the energy—dipole moments, transition probabilites, conductivities, etc.[2] The true wavefunctions are given by

$$\Psi_k = e^{-S} \widetilde{\Psi}_k = U_{kk} e^{-S} \Sigma_l \Phi_l C_{lk}, \tag{52}$$

---

1. The eigenvalues of a matrix $\boldsymbol{A}$ are solutions to the secular equation $det(\boldsymbol{A} - \lambda \boldsymbol{I}) = 0$. The determinant of a triangular matrix is equal to the product of the diagonal elements. (For example, see Theorem 4.1C of Ref. [10], pg.113.)

2. A similar situation occurs in density-functional theory, when Kohn-Sham wavefunctions are used for a many-electron system. This method is supposed to generate the correct energy for the system (for an exact density-functional) but does not determine the true wavefunctions [11]. In the present case, the true wavefunctions *are* known, i.e., Eq. (52).

which follows from Eq. (36). The $\Psi_k$ are orthogonal to one another, and the constant $U_{kk}$ is defined by requiring normalization to unity.

The above result suggests that physical considerations could also be useful in selecting the synthetic Hamiltonian. An example of that approach is given in Sec. 7.

## 5.5  Comparison with Variational Methods

It is interesting to compare the method discussed here with the variational method. In variational calculations, a trial function is selected, usually on the basis of physical arguments, and the energy is minimized with respect to the variational parameters.

Most calculations for many-electron systems use one-electron wavefunctions, which generate relatively simple equations but do not give a satisfactory treatment of electron correlation effects. In variational calculations, correlation effects are introduced by multiplying the one-electron wavefunctions by some function that depends explicitly on the electron-electron distances. However, integrals involving these correlated wavefunctions are quite complicated, and extensive numerical computation is generally involved.

The perturbation method has several potential advantages over the variational approach in such problems. The integrals involved are much less complicated. There is less guesswork in determining the function to be used. And the energy can be calculated exactly by carrying the iteration to convergence.

In fact, I have already used the concept of a synthetic Hamiltonian to examine the effects of electron correlation in the homogeneous electron gas [9]. In that case, the perturbation $W$ consists of the Coulomb repulsion between the electrons. The synthetic term $G$ was used to cancel the long-range part of the Coulomb repulsion, replacing it with a screened potential—an example of "Strategy #1," discussed above. The new iterative perturbation algorithm, Sec. 4, will allow extensions to that work.

# 6. HARMONIC OSCILLATOR WITH LINEAR PERTURBATION

Reed [8] has shown that the one-dimensional harmonic oscillator with a linear perturbation provides a simple but interesting test of perturbation theory. In this section, I will consider this problem using the three methods discussed above.

## 6.1 Statement of the Problem

First review the equations for the unperturbed oscillator [2]. The Hamiltonian is

$$\mathcal{H}_0 = -(\hbar^2/2m)\, d^2/dx^2 + (m\omega_0^2/2)x^2 = \tfrac{1}{2}(-d^2/d\xi^2 + \xi^2)\hbar\,\omega_0, \quad (53)$$

where $\xi = \sqrt{m\omega_0/\hbar}\,x$. In what follows, I will use units such that $\hbar\,\omega_0 = 1$.[1] Then the unperturbed energy levels are

$$E_n^0 = n + 1/2, \text{ for } n = 0, 1, 2, \ldots . \quad (54)$$

The normalized, orthogonal wavefunctions for the unperturbed oscillator are

$$\psi_n^0(\xi) = N_n H_n(\xi)e^{-\xi^2/2}, \, N_n = \pi^{-1/4}(2^n n!)^{-1/2}, \quad (55)$$

where $H_n(\xi)$ is a Hermite polynomial. The matrix elements of an operator $\Omega$ are

$$\langle n\,|\Omega|\,m\rangle = \int_{-\infty}^{\infty} \psi_n^0(\xi)\,\Omega\,\psi_m^0(\xi)\,d\xi. \quad (56)$$

For $\Omega = \xi$, the only non-zero matrix elements are those for which $m = n \pm 1$.

$$\langle n\,|\xi|\,n+1\rangle = \sqrt{(n+1)/2}, \langle n\,|\xi|\,n-1\rangle = \sqrt{n/2}. \quad (57)$$

(These relations, along with other matrix elements, are derived in Appendix C.)

Now consider a linear perturbation term, $W = \beta\xi$. The Hamiltonian for the perturbed oscillator can be written

$$\mathcal{H} = \mathcal{H}_0 + W = -\tfrac{1}{2}d^2/d\xi^2 + \tfrac{1}{2}\xi^2 + \beta\xi = -\tfrac{1}{2}d^2/d\mu^2 + \tfrac{1}{2}\mu^2 - \tfrac{1}{2}\beta^2, \quad (58)$$

where $\mu = \xi + \beta$. Hence the perturbed energy levels are

$$E_n = E_n^0 - \tfrac{1}{2}\beta^2 = n + \tfrac{1}{2} - \tfrac{1}{2}\beta^2 . \tag{59}$$

---

1. The definition of the problem given here differs slightly from that of Reed [8], who expressed energy in units of $\hbar\,\omega_0/2$.

## 6.2 Perturbed Oscillator with True Hamiltonian

The perturbation matrix elements are

$$W_{nm} = \langle n | W | m \rangle = \beta\delta_{m,\,n+1}\sqrt{(n+1)/2} + \beta\delta_{m,\,n-1}\sqrt{n/2} . \tag{60}$$

Substituting Eqs. (54) and (60) into the second-order expression for RSPT, Eq. (18), gives the exact energy—for all values of $n$ and $\beta$.

However, this apparent triumph of RSPT is misleading. The corresponding expression for the wavefunction, Eq. (17), does *not* give the exact result. When one attempts to go to higher order, numerical calculations show that the series fails to converge for large values of $n$ and that convergence is worst for large values of $\beta$.

The reason for this difficulty can be seen by inspecting the ratio

$$R_n = W_{n,\,n-1}/(E_n^0 - E_{n-1}^0) = \beta\sqrt{n/2} , \tag{61}$$

which appears in Eqs. (17) and (18). This ratio becomes larger, without bound, as $n$ increases. The numerical calculations show that convergence only occurs for values $R_n \lesssim 1$—as generally expected for RSPT.

The iterative method discussed in Sec. 3 does not give the exact energy at the first iteration, Eq. (32), but it does return the correct value after many iterations—for those eigenstates for which it converges. It gives generally better convergence than RSPT (as well as being much faster in numerical calculations.) With $\beta = 1/2$, for example, RSPT fails to converge for $n > 11$, while the iterative method converges for $n \leq 20$. On the other hand, the iterative scheme gives poorer results for large values of $\beta$, failing to converge at all for $\beta > 1$.

## 6.3 Perturbed Oscillator with Synthetic Hamiltonian

As noted in Sec. 5, one generates a synthetic Hamiltonian $\widetilde{\mathcal{H}}$ by choosing a function $S$ and applying Eqs. (43), (44), and (45). Since $\widetilde{\mathcal{H}}$ has the same energy eigenvalues as $\mathcal{H}$, the goal is to find a function $S$ that gives good results when perturbation theory is applied.

For the present problem, consider the function

$$S = A\xi , \tag{62}$$

where the constant $A$ can be chosen at will. *It is important to note that $A$ is neither a fitting parameter nor a variational parameter*, since *any* value will give the same eigenvalues—provided the series converges.

It is readily seen that

$$dS/d\xi \;=\; A \;\text{ and }\; G \;=\; \tfrac{1}{2}(dS/d\xi)^2 \;=\; \tfrac{1}{2}A^2 . \tag{63}$$

Since $W$ and $\mathcal{F}$ have no diagonal elements, the diagonal matrix elements of $\widetilde{\boldsymbol{H}}$ are

$$\widetilde{H}_{nn} \;=\; \langle n \,|\, \mathcal{H}_0 - G \,|\, n \rangle \;=\; n + \tfrac{1}{2} - \tfrac{1}{2}A^2 . \tag{64}$$

The off-diagonal matrix elements are

$$\widetilde{H}_{nm} \;=\; \langle n \,|\, W + \mathcal{F} \,|\, m \rangle \;=\; [\beta + (E_m^0 - E_n^0)A]\langle n \,|\xi|\, m \rangle, \; m = n\pm 1 . \tag{65}$$

Now employ "Strategy #2," as discussed in Sec. 5.3: Choosing $A = \pm\beta$ makes $\widetilde{\boldsymbol{H}}$ into a triangular matrix. (The + sign zeros out the lower triangle, $m < n$, while the - sign zeros out the upper triangle, $m > n$.) In either case, the diagonal elements of $\widetilde{\boldsymbol{H}}$ are also the eigenvalues, giving

$$E_n \;=\; \widetilde{E}_n \;=\; \widetilde{H}_{nn} \;=\; n + \tfrac{1}{2} - \tfrac{1}{2}\beta^2 , \tag{66}$$

in agreement with Eq. (59).

The above analysis gives the correct eigenvalues without using either RSPT or the iterative method. However, perturbation theory must still be used to compute the wavefunctions. When that is done, numerical calculations using both methods give full convergence—for *all* values of $n$ and $\beta$.

Of course, it will not be possible to completely triangularize $\widetilde{\boldsymbol{H}}$ in most cases. However, partial triangularization would still be expected to improve convergence. To test this assertion, I tried perturbation calculations using $A = \beta/2$. Numerical calculations still gave the correct answer in this case, even though convergence was not as good.

# 7. HARMONIC OSCILLATOR WITH QUARTIC PERTURBATION

Calculating the energy levels of a harmonic oscillator with a quartic potential term is a difficult problem for perturbation theory. The Hamiltonian is

$$\mathcal{H} = -\tfrac{1}{2}d^2/d\xi^2 + \tfrac{1}{2}\xi^2 + \beta\xi^4 = \mathcal{H}_0 + W. \tag{67}$$

## 7.1 Standard Approach

Bender and Wu [12] have shown that RSPT diverges for *any* value of the parameter $\beta$. The reason for this fact is that the matrix elements of $\xi^4$ are on the order of $n^2$ and so become very large for highly excited states. (See Appendix C). Lynch and Mavromatis [13] showed that reasonable results for the ground state can be obtained by using RSPT with a finite number of excited states, i.e., by truncating the Hamiltonian matrix. But the series eventually fails to converge as the size of the matrix is increased.

My own numerical calculations agree with those observations for both RSPT and the iterative perturbation method. For $\beta = 1$, both methods diverge for matrices larger than 7 x 7. For a 7 x 7 matrix, both methods converge for the ground state, giving an energy slightly higher than the exact value.

## 7.2 Synthetic Hamiltonian

Some of the above difficulties can be overcome by using a synthetic Hamiltonian.

Insight into the nature of the problem can be gained by looking at the asymptotic behavior as $\xi \rightarrow \pm\infty$. The Schrödinger equation gives

$$d^2\psi/d\xi^2 = 2\beta\xi^4\psi, \ \psi \rightarrow \exp(\pm\sqrt{2\beta}\xi^3/3) \text{ as } \xi \rightarrow \pm\infty. \tag{68}$$

In order to have $\psi \rightarrow 0$ as $\xi \rightarrow \pm\infty$, one must take the - sign for $\xi > 0$ and the + sign for $\xi < 0$. This fact suggests that an approximate solution to the wavefunction might have the form

$$\psi \approx e^{-S}\psi_n^0, \text{ where } S = \sqrt{2\beta}\left|\xi^3\right|/3. \tag{69}$$

Given this insight, let us generate a synthetic Hamiltonian $\widetilde{\mathcal{H}}$ using the function

$$S(\xi) = A_2\xi^2 + A_3\lambda\xi^3, \tag{70}$$

where $A_3 = \sqrt{2\beta}/3$ and $\lambda = |\xi|/\xi$.

Once again, note that *the constants $A_2$ and $A_3$ are not fitting parameters or variational parameters*. They are chosen to give a Hamiltonian matrix that is suitable for perturbation theory.

After computing the function $G$ from Eq. (45), one obtains

$$W - G = -2A_2^2 \xi^2 - 6A_2 A_3 \lambda \xi^3. \tag{71}$$

Hence this choice of $S$ cancels the $\xi^4$ perturbation term and introduces terms of order $\xi^2$ and $\xi^3$ —a variation of "Strategy #1," discussed in Sec. 5.3. The matrix elements of $\tilde{\boldsymbol{H}}$ are

$$\begin{aligned}\widetilde{H}_{nm} =\ & (n + \tfrac{1}{2})\delta_{nm} - (2A_2 + n - m)A_2 \langle n\,|\xi^2|\,m\rangle \\ & - (6A_2 + n - m)A_3 \langle n\,|\lambda\xi^3|\,m\rangle.\end{aligned} \tag{72}$$

If one sets $A_2 = 0$, $W - G = 0$, and all the off-diagonal terms come from operator $\mathcal{F}$, as discussed in Sec. 5.3, Strategy #1. However, calculations with $A_2 = 0$ do not converge, due to the fact that the matrix elements $\langle n\,|\lambda\xi^3|\,m\rangle$ are so large.

As shown in Appendix C, $\langle n\,|\lambda\xi^3|\,m\rangle$ vanishes when $n + m$ is odd but is nonzero when $n + m$ is even, the largest values occurring when $m = n$ and $m = n \pm 2$. The contributions from these terms can be reduced by making the appropriate choice for $A_2$. As a rough estimate, one could choose $A_2 = -1/3$,[1] which sets the $m = n - 2$ term to zero in the lower triangle of the matrix. A slightly better result can be obtained by considering the quantity

$$Q_n = \Sigma_{m \neq n}(6A_2 + n - m)(6A_2 - n + m)\langle n\,|\lambda\xi^3|\,m\rangle^2, \tag{73}$$

which is obtained by taking the product of the $\langle n\,|\lambda\xi^3|\,m\rangle$ matrix elements in the upper and lower triangles and summing over all terms. Choosing the value of $A_2$ that sets $Q_n = 0$ gives

$$A_2^2 = \frac{\Sigma_{m \neq n}(m - n)^2 \langle n\,|\lambda\xi^3|\,m\rangle^2}{36\Sigma_{m \neq n}\langle n\,|\lambda\xi^3|\,m\rangle^2}. \tag{74}$$

---

1. A positive sign for $A_2$ would set the corresponding term to zero in the upper triangle. However, the negative value is chosen because calculations using the positive value do not converge. The reason for this fact can be understood by inspecting Eq. (71). The quadratic and cubic terms tend to cancel one another when $A_2$ and $A_3$ have opposite signs.

The value of $A_2$ obtained from Eq. (74) depends on the quantum number $n$, ranging from $A_2 = -0.375$ for $n = 0$, to $A_2 = -0.344$ for $n \geq 10$. The fact that these values are so close to -1/3, the rough estimate made above, shows that the $m = n + 2$ term makes the largest contribution in Eq. (74).

Numerical calculations using the synthetic Hamiltonian give far better convergence than those with the true Hamiltonian, allowing calculation of the energies for several excited states as well as the energy of the ground state. The results of calculations using the iterative perturbation method are shown in Fig. 1 and Table 1. These calculations were made using a 100 x 100 Hamiltonian matrix, with $A_2 = -0.35$ for $\beta \leq 0.5$ and $A_2 = -0.375$ for $\beta \geq 0.6$.

Figure 1 shows the energies of the first six levels of the perturbed harmonic oscillator as functions of the perturbation parameter $\beta$. Figure 1a compares the calculated results for the ground state ($n = 0$) with those presented in Table II (labeled "exact values") of Ref. [13], shown as circles. The agreement is perfect except at $\beta = 0.25$, where the value given in [13] is clearly in error because it is inconsistent with the other entries in the table.

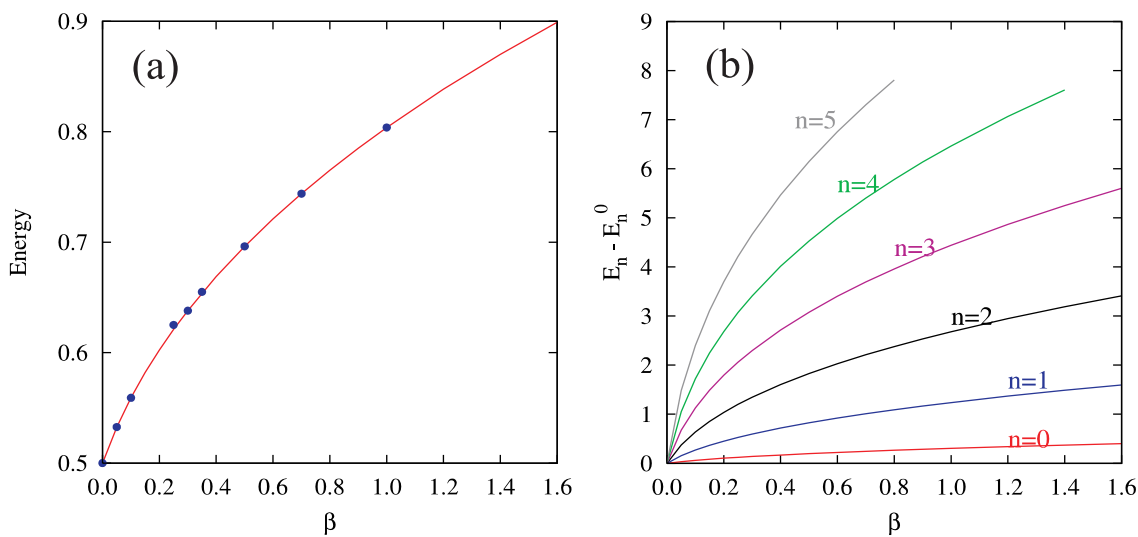

**Fig. 1. Energy levels of harmonic oscillator with quartic perturbation. (a) Ground-state energy vs. perturbation parameter $\beta$. Solid line is calculated from perturbation theory, circles are values from Ref. [13]. (b) Corrections to energies of first six states vs. $\beta$, calculated from perturbation theory.**

Figure 1b shows the perturbation correction to the energy, $E_n - E_n^0$ as a function of $\beta$ for the lowest lying energy levels. These results are vastly superior to those using the true Hamiltonian, where the ground state energy failed to converge for a matrix larger than 7 x 7.

RSPT does not give as good convergence as the iterative method and requires more computing time. The shaded blocks in Table 1 show the values of β and $n$ where RSPT failed to converge. It can be seen that the iterative method converges over a significantly wider range of parameters, particularly for large values of β. However, the RSPT energies are identical to those for the iterative method in those cases where convergence does occur.

**Table 1: Energy $E_n^0$ of a harmonic oscillator with a quartic perturbation. Calculations were made with a synthetic Hamiltonian, the iterative perturbation method, and a 100 x 100 Hamiltonian matrix. Results are shown only for states that converged. Shaded blocks show cases where RSPT failed to converge.**

| β | $n = 0$ | $n = 1$ | $n = 2$ | $n = 3$ | $n = 4$ | $n = 5$ | $n = 6$ | $n = 7$ |
|---|---------|---------|---------|---------|---------|---------|---------|---------|
| 0.00 | 0.50000 | 1.5000 | 2.5000 | 3.5000 | 4.5000 | 5.5000 | 6.5000 | 7.5000 |
| 0.05 | 0.53264 | 1.6534 | 2.8740 | 4.1763 | 5.5493 | 6.9850 | 8.4774 | 10.022 |
| 0.10 | 0.55915 | 1.7695 | 3.1386 | 4.6289 | 6.2203 | 7.8998 | 9.6578 | 11.487 |
| 0.15 | 0.58202 | 1.8662 | 3.3529 | 4.9877 | 6.7444 | 8.6065 | 10.562 | 12.602 |
| 0.20 | 0.60240 | 1.9505 | 3.5363 | 5.2913 | 7.1845 | 9.1963 | 11.313 | 13.525 |
| 0.25 | 0.62093 | 2.0260 | 3.6985 | 5.5576 | 7.5684 | 9.7091 | 11.965 | 14.323 |
| 0.30 | 0.63799 | 2.0946 | 3.8448 | 5.7966 | 7.9118 | 10.166 | 12.544 | 15.033 |
| 0.40 | 0.66877 | 2.2169 | 4.1028 | 6.2156 | 8.5114 | 10.963 | 13.552 | – |
| 0.50 | 0.69617 | 2.3244 | 4.3275 | 6.5784 | 9.0288 | 11.649 | 14.418 | – |
| 0.60 | 0.72104 | 2.4210 | 4.5281 | 6.9011 | 9.4877 | 12.256 | – | – |
| 0.70 | 0.74390 | 2.5092 | 4.7103 | 7.1933 | 9.9026 | – | – | – |
| 0.80 | 0.76514 | 2.5907 | 4.8779 | 7.4614 | 10.283 | – | – | – |
| 0.90 | 0.78503 | 2.6666 | 5.0336 | 7.7101 | 10.635 | – | – | – |
| 1.00 | 0.80377 | 2.7379 | 5.1793 | 7.9424 | – | – | – | – |
| 1.20 | 0.83840 | 2.8690 | 5.4464 | 8.3675 | – | – | – | – |
| 1.40 | 0.86996 | 2.9878 | 5.6876 | 8.7508 | – | – | – | – |
| 1.60 | 0.89907 | 3.0969 | 5.9085 | – | – | – | – | – |

I also performed calculations for 125 x 125 and 150 x 150 matrices, to investigate the effect of matrix size. Convergence was poorer for the larger matrices. The energies of the converged states did not vary, over the range studied, but the maximum values of β and $n$ that yielded converged solutions did decrease with size. It is not clear if this lack of convergence for large matrices is a limitation of the method or if roundoff errors and other numerical problems contribute to the size effect.

It is also likely that better convergence could be obtained by improving on the function $S$ used to generate the synthetic Hamiltonian.

# 8. TWO-DIMENSIONAL OSCILLATOR WITH PERTURBATION

In this section, I will apply the perturbation techniques to a two-dimensional oscillator. The Hamiltonian is

$$\mathcal{H} = h_0(\xi_1) + h_0(\xi_2) + W(\xi_1, \xi_2) = \mathcal{H}_0 + W, \text{ where} \tag{75}$$

$$h_0(\xi) = -\tfrac{1}{2}d^2/d\xi^2 + \tfrac{1}{2}\xi^2 \text{ and } W = \beta\xi_1\xi_2. \tag{76}$$

The energies levels of the unperturbed system are

$$E^0_{n_1 n_2} = n_1 + n_2 + 1. \tag{77}$$

Hence the excited states of the unperturbed system are degenerate. The perturbation breaks this degeneracy, as shown below.

## 8.1 Exact Solution

An exact solution for the eigenvalues can be obtained using the transformation

$$\xi_1 = (\alpha_1\mu_1 + \alpha_2\mu_2)/\sqrt{2} , \xi_2 = (\alpha_1\mu_1 - \alpha_2\mu_2)/\sqrt{2} , \text{ where} \tag{78}$$

$$\alpha_1 = (1+\beta)^{-1/4} \text{ and } \alpha_2 = (1-\beta)^{-1/4}. \tag{79}$$

Substituting these equations into Eq. (75) gives

$$\mathcal{H} = \sqrt{1+\beta}\, h_0(\mu_1) + \sqrt{1-\beta}\, h_0(\mu_2). \tag{80}$$

Hence the energy levels of the perturbed oscillator are

$$E_{n_1 n_2} = \sqrt{1+\beta}\, (n_1 + \tfrac{1}{2}) + \sqrt{1-\beta}\, (n_2 + \tfrac{1}{2}) \text{ (for } \beta \leq 1). \tag{81}$$

## 8.2 Perturbation Calculations

The iterative perturbation method of Sec. 4 is capable of finding solutions for those cases where the unperturbed state is non-degenerate or doubly-degenerate. In this example, it can also find some of the solutions for states of higher degeneracy, as shown below.

Direct application of the method to the true Hamiltonian gives fairly good results. However, better convergence can be obtained using a synthetic Hamiltonian. In this example, I will use the function

$$S = A\xi_1\xi_2, \tag{82}$$

which gives the following matrix elements.

$$
\begin{aligned}
\langle n_1 n_2 \mid \widetilde{\mathcal{H}} \mid m_1 m_2 \rangle = {} & (n_1 + n_2 + 1)\delta_{n_1 m_1}\delta_{n_2 m_2} \\
& + [\beta + (m_1 + m_2 - n_1 - n_2)A]\langle n_1 \mid \xi \mid m_1\rangle\langle n_2 \mid \xi \mid m_2\rangle \\
& - \tfrac{1}{2}A^2[\langle n_1 \mid \xi^2 \mid m_1\rangle\delta_{n_2 m_2} + \langle n_2 \mid \xi^2 \mid m_2\rangle\delta_{n_1 m_1}].
\end{aligned}
\tag{83}
$$

Since the elements $\langle n \mid \xi \mid m \rangle$ vanish except for $m = n \pm 1$, choosing $A = \beta / 2$ eliminates the second term on the right side of Eq. (83) for all elements having $m_1 < n_1$ and $m_2 < n_2$. This choice was used in the calculations discussed below.

Figure 2 shows the first six energy levels of the perturbed 2-D oscillator, as functions of $\beta$, over the range $0 \le \beta \le 1$. The exact results, Eq. (81), are shown as solid lines, the perturbation results as discrete points. The perturbation calculations were made with a Hamiltonian matrix with quantum numbers $0 \le n_1 \le 39$ and $0 \le n_2 \le 39$, i.e., a matrix of dimensions 820 x 820.

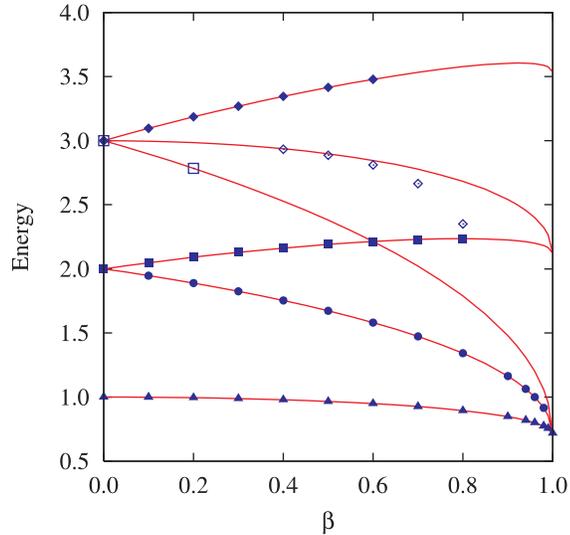

As expected, the iterative method found solutions for the ground state and the doubly-degenerate first excited state of the unperturbed oscillator. It also found solutions for some higher energy states of three- and four-fold degeneracy. The perturbation results agree with the exact results for the lowest three levels. The method did *not* always give the correct result for the higher levels, like those shown by the open diamonds; I have not yet determined the reason for this discrepancy.

**Fig. 2. Energy levels of perturbed 2-D oscillator as functions of $\beta$. Solid lines are exact solutions; discrete points are calculations using iterative perturbation method and synthetic Hamiltonian.**

The success of the iterative method in dealing with cases of degeneracy demonstrates its value as an alternative to the standard RSPT method. Further improvements to the theory could make this approach even more useful.

# 9. SUMMARY AND CONCLUSIONS

This report takes a fresh look at the use of perturbation theory for solving time-independent problems in quantum mechanics. It presents two new ideas—new ways to derive and solve the perturbation equations, and new ways to construct the perturbation matrix elements used in the solutions.

The new solution methods are discussed in Secs. 2-4. The basic equations are first written in a way that is closely related to the matrix diagonalization method. Two perturbation algorithms are then derived. The first is the Raleigh-Schrödinger method (RSPT) with new options for treating diagonal elements. The other algorithm uses a completely different method for solving the equations, an iterative technique that does not use a power series expansion. This iterative method can also be used in cases of degeneracy.

The second innovation in this report is the introduction of "synthetic" Hamiltonians. The true Hamiltonian for the system is replaced with a modified one that has the same eigenvalues. This approach allows one to cancel part of the perturbation term and to reduce the magnitude of the off-diagonal matrix elements.

These techniques have been illustrated in three examples involving perturbed harmonic oscillators.

*   For a 1-D oscillator with a linear perturbation, the iterative method gives better convergence than RSPT when applied to the true Hamiltonian matrix. When a synthetic Hamiltonian is used, the eigenvalues can be obtained without *any* perturbation expansion. And the perturbation expansions, when used to compute the wavefunctions, converge for all values of the parameter, in contrast to the case of the true Hamiltonian.

*   For a 1-D oscillator with a quartic perturbation, neither method converges when the true Hamiltonian is used—as others have previously shown. But, with a synthethic Hamiltonian, it is possible to obtain exact solutions for the ground state and several excited states, over a wide range of parameters, without extensive numerical work.

*   For a perturbed 2-D oscillator, the iterative method gives correct solutions for the non-degenerate ground state and the doubly-degenerate first exited state of the unperturbed oscillator. It also finds solutions for some of the states of higher degeneracy. Here again, a synthetic Hamiltonian gives better convergence than the true Hamiltonian.

The new concepts presented here do have some limitations and will undoubtedly require further work before they can be fully exploited. Further development and refinement of these ideas offers considerable potential for future applications.

# Appendix A

# Listing of RSPT Subroutine

```
      SUBROUTINE QRSPT(H,N,MX,ERE,ERC,E,C,NIT,CK)
C----------------------------------------------------------------------
C
C  SUBROUTINE.   QRSPT(H,N,MX,ERE,ERC,E,C,NIT,CK)
C
C  PURPOSE.      Calculate eigenvalues and eigenvectors of N X N
C                Hamiltonian matrix by Raleigh-Schrodinger theory
C
C  ARGUMENTS.    H    (input) - Hamiltonian matrix
C                N    (input) - number of elements in H matrix
C                MX   (input) - dimension of 1st index of H & C arrays
C                               as specified in calling routine
C                ERE  (input) - error criterion for energies
C                ERC  (input) - error criterion for coefficients
C                E    (output) - vector of eigenvalues
C                C    (output) - coefficent matrix
C                NIT  (output) - matrix giving number of iterations for
C                               each eigenvalue, or NIT=0 when maximum
C                               number of iterations is exceeded
C                CK   (input) - scratch array, dimensioned (MX,1000)
C
C  REMARKS.      C is matrix that diagonalizes H:
C                --If H is Hermitian, transformation is unitary,
C                  C**(-1) = CT (C-transpose)
C                --If H is not Hermitian, transformation is not
C                  unitary, and CT is not inverse of C.
C                Iterations stop when relative errors in energy and C
C                  drop below specified values or when IMX iterations
C                  have been exceeded.
C                Note that each eigenvalue is computed independently.
C                  Failure of one does not imply failure of all.
C                C coefficients normalize wavefunctions to unity.
C                Energies & coefficients for different orders are saved
C                  in arrays EK & CK, The highest order being 1000.
C
C  EXTERNALS.    None
C
C  PROGRAMMER.   G. I. Kerley
C
C  DATE.         28 November 2006
C
C----------------------------------------------------------------------
      IMPLICIT DOUBLE PRECISION (A-H,O-Z)
      SAVE
      LOGICAL LOOP
      PARAMETER (ZRO=0.0D0,ONE=1.0D0,TWO=2.0D0,FOR=4.0D0,IMX=1000)
```

```
        DIMENSION H(MX,*),E(*),C(MX,*),NIT(*)
        DIMENSION CK(MX,*),EK(1000)
* Loop over all eigenstates
        DO 9 K=1,N
* Initialize E & C, EK & CK arrays
        EK(1) = ZRO
        CK(K,1) = ZRO
        C(K,K) = ONE
        DO 1 L=1,N
         IF(K.NE.L) THEN
          CK(L,1) = H(L,K)/(H(K,K)-H(L,L))
          EK(1) = EK(1)+H(K,L)*CK(L,1)
          C(L,K) = CK(L,1)
         ENDIF
         E(K) = H(K,K)+EK(1)
 1      CONTINUE
* Iteration loop--compute coefficients and energy for i-th order
        DO 5 I=2,IMX
        LOOP = .FALSE.
        NIT(K) = I
* Compute new values of EK & CK
        EK(I) = ZRO
        DO 4 L=1,N
         CK(L,I) = ZRO
         IF(L.NE.K) THEN
          DO 2 J=1,N
           IF(J.NE.L .AND. J.NE.K)
     &       CK(L,I) = CK(L,I)+H(L,J)*CK(J,I-1)
 2        CONTINUE
          DO 3 J=1,I-1
           CK(L,I) = CK(L,I)-CK(L,I-J)*EK(J-1)
 3        CONTINUE
          CK(L,I) = CK(L,I)/(H(K,K)-H(L,L))
          EK(I) = EK(I)+H(K,L)*CK(L,I)
          C(L,K) = C(L,K)+CK(L,I)
          CTST = ABS(CK(L,I))-ERC*ABS(C(L,K))
          IF(CTST.GT.ZRO) LOOP=.TRUE.
         ENDIF
 4      CONTINUE
        E(K) = E(K)+EK(I)
        ETST = ABS(EK(I))-ERE*ABS(E(K))
        IF(ETST.GT.ZRO) LOOP=.TRUE.
        IF(.NOT.LOOP) GO TO 6
 5      CONTINUE
        NIT(K) = 0
 6      CONTINUE
* Normalize C coefficients for this eigenstate
        UKK = ZRO
        DO 7 L=1,N
         UKK = UKK+C(L,K)**2
 7      CONTINUE
        UKK = ONE/SQRT(UKK)
        DO 8 L=1,N
         C(L,K) = UKK*C(L,K)
```

```
8    CONTINUE
9   CONTINUE
    RETURN
    END
```

# Appendix B

## Listing of Iterative Perturbation Subroutine

```
      SUBROUTINE QIPRT(H,N,MX,ERE,ERC,E,C,NIT)
C----------------------------------------------------------------------
C
C  SUBROUTINE.   QIPRT(H,N,MX,ERE,ERC,E,C,NIT)
C
C  PURPOSE.      Compute eigenvalues and eigenvectors of N X N
C                Hamiltonian matrix by iterative perturbation method.
C
C  ARGUMENTS.    H    (input) - Hamiltonian matrix
C                N    (input) - number of elements in H matrix
C                MX   (input) - dimension of 1st index of H & C arrays
C                               as specified in calling routine
C                ERE  (input) - error criterion for energies
C                ERC  (input) - error criterion for coefficients
C                E    (output) - vector of eigenvalues
C                C    (output) - coefficent matrix, dimension (MX,N+1)
C                NIT (output) - matrix giving number of iterations for
C                               each eigenvalue, or NIT=0 when maximum
C                               number of iterations is exceeded
C
C  REMARKS.      C is matrix that diagonalizes H:
C                --If H is Hermitian, transformation is unitary,
C                  C**(-1) = CT (C-transpose)
C                --If H is not Hermitian, transformation is not
C                  unitary, and CT is not inverse of C.
C                Iterations stop when relative errors in energy and C
C                  drop below specified values or when IMX iterations
C                  have been exceeded.
C                Note that each eigenvalue is computed independently.
C                  Failure of one does not imply failure of all.
C                C coefficients normalize wavefunctions to unity.
C
C  EXTERNALS.    None
C
C  PROGRAMMER.   G. I. Kerley
C
C  DATE.         28 November 2006
C
C----------------------------------------------------------------------
      IMPLICIT DOUBLE PRECISION (A-H,O-Z)
      SAVE
      LOGICAL LOOP
      PARAMETER (ZRO=0.0D0,ONE=1.0D0,TWO=2.0D0,FOR=4.0D0,IMX=10000,
     &           ERR=1.0D-20)
      DIMENSION H(MX,*),E(*),C(MX,*),NIT(*)
* Loop over all eigenstates
```

```
      DO 9 K=1,N
* Initialize E & C arrays
       E(K) = H(K,K)
       DO 1 L=1,N
        C(L,K) = ZRO
 1     CONTINUE
* Iteration loop--update quantities & compute energy.
       DO 5 I=1,IMX
        LOOP = .FALSE.
        NIT(K) = I
* Compute new values of C
        DO 3 L=1,N
         IF(L.NE.K) THEN
          DKL = H(K,K)-H(L,L)
          IF(ABS(DKL).GT.ERR) THEN
           SKL = SIGN(ONE,DKL)
          ELSE
           SKL = SIGN(1,K-L)
          ENDIF
          WLK = H(L,K)
          DO 2 J=1,N
           IF(J.NE.L .AND. J.NE.K)
     &       WLK = WLK+(H(L,J)-C(L,K)*H(K,J))*C(J,K)
 2        CONTINUE
          QLK = DKL**2+FOR*H(K,L)*WLK
          IF(QLK.GE.ZRO) THEN
           QLK = (SQRT(QLK)+ABS(DKL))/TWO
           IF(QLK.NE.ZRO) THEN
            C(L,N+1) = SKL*WLK/QLK
           ELSE
            C(L,N+1) = SKL
           ENDIF
          ELSEIF(H(K,L).NE.ZRO) THEN
           C(L,N+1) = -DKL/TWO/H(K,L)
          ELSE
           NIT(K) = 0
           GO TO 6
          ENDIF
         ENDIF
 3      CONTINUE
* Update C array and compute energy
        ESAV = E(K)
        E(K) = H(K,K)
        DO 4 L=1,N
         IF(L.NE.K) THEN
          CTST = ABS(C(L,K)-C(L,N+1))-ERC*ABS(C(L,K)+C(L,N+1))/TWO
          IF(CTST.GT.ZRO) LOOP=.TRUE.
          C(L,K) = C(L,N+1)
          E(K) = E(K)+H(K,L)*C(L,K)
         ENDIF
 4      CONTINUE
        ETST = ABS(E(K)-ESAV)-ERE*ABS(E(K)+ESAV)/TWO
        IF(ETST.GT.ZRO) LOOP=.TRUE.
        IF(.NOT.LOOP) GO TO 6
```

```
5      CONTINUE
       NIT(K) = 0
6      CONTINUE
* Normalize C coefficients for this eigenstate
       C(K,K) = ONE
       UKK = ZRO
       DO 7 L=1,N
        UKK = UKK+C(L,K)**2
7      CONTINUE
       UKK = ONE/SQRT(UKK)
       DO 8 L=1,N
        C(L,K) = UKK*C(L,K)
8      CONTINUE
9     CONTINUE
      RETURN
      END
```

# Appendix C

# Calculation of Harmonic Oscillator Matrix Elements

This appendix derives formulas for the matrix elements of $\xi$, $\xi^2$, $\xi^3$, $\xi^4$, $\lambda\xi$, and $\lambda\xi^3$, where $\lambda = |\xi|/\xi$.

The Hermite polynomials satisfy the well-known recursion formula [2]

$$H_{n+1}(\xi) = 2\xi H_n(\xi) - 2n H_{n-1}(\xi) \text{ (with } H_0 = 1 \text{)}. \tag{C.1}$$

Using this equation, together with Eq. (55), one finds the following formula for the harmonic oscillator wavefunctions.

$$\xi\psi_n^0(\xi) = \sqrt{(n+1)/2}\ \psi_{n+1}^0(\xi) + \sqrt{n/2}\ \psi_{n-1}^0(\xi). \tag{C.2}$$

Now multiply the above equation by $\psi_m^0$ and integrate. Using the orthonormality of the wavefunctions,

$$\langle m\,|\xi|\,n\rangle = \langle n\,|\xi|\,m\rangle = \delta_{m,\,n+1}\sqrt{(n+1)/2} + \delta_{m,\,n-1}\sqrt{n/2}, \tag{C.3}$$

in agreement with Eq. (57).

Note that the second term in Eq. (C.3) can be obtained from the first term by interchanging the values of $n$ and $m$. In what follows, I will only give the results for $m \geq n$; those for $m < n$ can be obtained by the same kind of interchange.

The equations for $\xi^2$, $\xi^3$, and $\xi^4$ can be derived by applying the sum rule

$$\langle n\,|\xi^{i+j}|\,m\rangle = \Sigma_k\,\langle n\,|\xi^i|\,k\rangle\langle k\,|\xi^j|\,m\rangle. \tag{C.4}$$

The resulting formulas (for $m \geq n$) are

$$\langle n\,|\xi^2|\,n\rangle = n + \tfrac{1}{2}, \tag{C.5}$$

$$\langle n\,|\xi^2|\,n+2\rangle = \tfrac{1}{2}\sqrt{(n+1)(n+2)}, \tag{C.6}$$

$$\langle n\,|\xi^3|\,n+1\rangle = \tfrac{3}{2}(n+1)\sqrt{(n+1)/2}, \tag{C.7}$$

$$\langle n \,|\xi^3|\, n+3 \rangle \;=\; \tfrac{1}{2}\sqrt{(n+1)(n+2)(n+3)/2}\,, \qquad (C.8)$$

$$\langle n \,|\xi^4|\, n \rangle \;=\; \tfrac{3}{4}(2n^2 + 2n + 1)\,, \qquad (C.9)$$

$$\langle n \,|\xi^4|\, n+2 \rangle \;=\; (n+\tfrac{3}{2})\sqrt{(n+1)(n+2)}\,, \qquad (C.10)$$

$$\langle n \,|\xi^4|\, n+4 \rangle \;=\; \tfrac{1}{4}\sqrt{(n+1)(n+2)(n+3)(n+4)}\,. \qquad (C.11)$$

The matrix elements of $\lambda\xi$ and $\lambda\xi^3$ cannot be obtained in this simple manner. Noting that $H_n(-\xi) = (-1)^n H_n(\xi)$, one finds

$$\langle n \,|\lambda\xi|\, m \rangle \;=\; N_n N_m [\, 1 + (-1)^{n+m}\,]\int_0^\infty H_n(\xi) H_m(\xi)\; e^{-\xi^2}\xi\, d\xi\,, \qquad (C.12)$$

$$\langle n \,|\lambda\xi^3|\, m \rangle \;=\; N_n N_m [\, 1 + (-1)^{n+m}\,]\int_0^\infty H_n(\xi) H_m(\xi)\; e^{-\xi^2}\xi^3\, d\xi\,. \qquad (C.13)$$

Hence these matrix elements have non-zero values only when is $n + m$ even, i.e., $m = n, n \pm 2, n \pm 4, n \pm 6, \dots$.

It is possible to express $H_n$ as an n-th order polynomial in $\xi$ and to calculate the coefficients from a recursion formula [2]. In that case, the integrals can be expressed as a double summation, and each term can be evaluated exactly. In numerical calculations, however, this procedure gives poor results for large values of $n$; the reason is that the polynomial coefficients are large and alternate in sign, giving large roundoff errors.

In the present work, I have obtained somewhat better results using numerical integration, calculating the Hermite polynomials from Eq. (C.1) instead of the polynomial formula. This method gives accurate results for matrix elements for which $n \lesssim 100$. Fortunately, good estimates for larger values of the quantum numbers can be made by extrapolation, as discussed below.

The dependence of the matrix elements $\langle n \,|\lambda\xi^3|\, m \rangle$ on the quantum numbers is shown in Fig. 3.

Figure 3a shows $\langle n \,|\lambda\xi^3|\, m \rangle$ as a function of $n$ for several values of $m$. For large values of $n$, the matrix elements are roughly proportional to $n^{3/2}$, for all values of $m$. They are largest for $m = n$ and $m = n + 2$, dropping off rapidly as $m$ increases. The matrix elements $\langle n \,|\lambda\xi|\, m \rangle$ show similar behavior, varying as $n^{1/2}$ for large $n$.

Figure 3b shows $\langle n \,|\lambda\xi^3|\, n+k \rangle$ as a function of $k$ for several values of $n$. The matrix elements have been multiplied by the factor $(k+1)^{5/2}$ to emphasize the behavior for large $k$. As already noted, the values are largest for $k = 0$ and $k = 2$, dropping off rapidly as $k$ increases. For large values of $k$, the matrix elements fall off as $k^{-5/2}$, alternating in sign. The matrix elements $\langle n \,|\lambda\xi|\, n+k \rangle$ show similar behavior, falling off as $k^{-5/4}$ for large $k$.

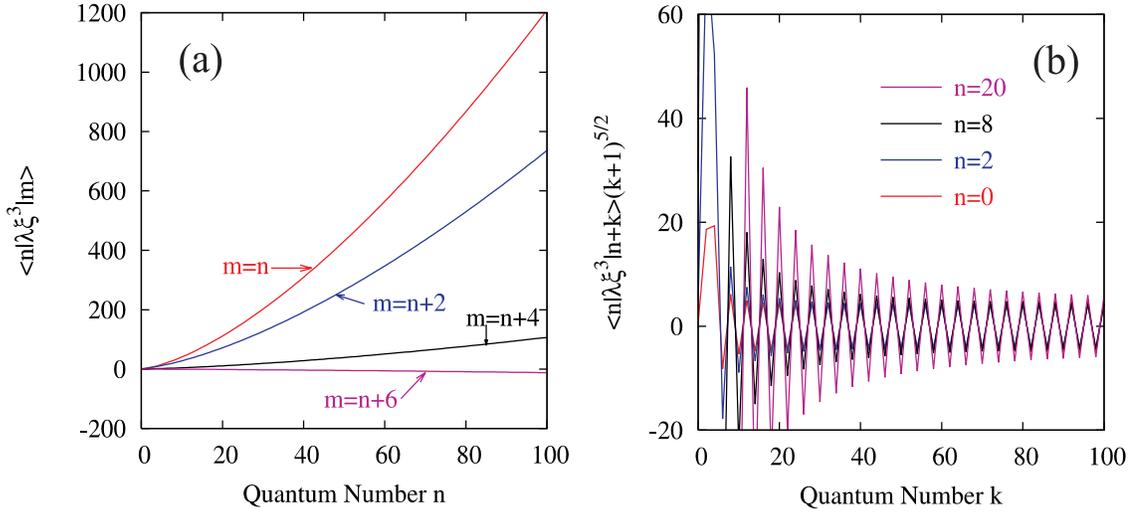

**Fig. 3. Dependence of matrix elements on quantum numbers. (a):** $\langle n|\lambda\xi^3|\,m \rangle$ **plotted as a function of $n$ for several values of $m$. b):** $\langle n \,|\lambda\xi^3|\, n+k \rangle(k+1)^{5/2}$ **plotted as a function of $k$ for several values of $n$.**

The following extrapolation procedure was used to calculate the matrix elements for large values of the quantum numbers. For (even) values of $k > k_e = 50$,

$$\langle n \,|\lambda\xi|\, n+k \rangle = (-1)^{(k_e+k)/2}\langle n \,|\lambda\xi|\, n+k_e \rangle(k_e/k)^{5/4}, \qquad \text{(C.14)}$$

$$\langle n \,|\lambda\xi^3|\, n+k \rangle = (-1)^{(k_e+k)/2}\langle n \,|\lambda\xi^3|\, n+k_e \rangle(k_e/k)^{5/2+0.02n}. \qquad \text{(C.15)}$$

For $n > n_e = 100 - k/2$ (where $k \le k_e$),

$$\langle n \,|\lambda\xi|\, n+k \rangle = \langle n_e \,|\lambda\xi|\, n_e+k \rangle(n/n_e)^{1/2}, \qquad \text{(C.16)}$$

$$\langle n \,|\lambda\xi^3|\, n+k \rangle = \langle n_e \,|\lambda\xi^3|\, n_e+k \rangle(n/n_e)^{3/2}. \qquad \text{(C.17)}$$

This procedure was used to generate tables to a maximum size of 500 x 500.